\documentstyle[epsfig]{aipproc}

\begin{document}
\thispagestyle{empty}

\begin{flushright}
DESY--97--109\\
June 1997
\end{flushright}

\vspace{2.5cm}

\begin{center}
{\huge\bf Spin Physics}
\vskip 1.5cm
{\large  T.~Gehrmann} 
\vskip .4cm
{\it DESY, Theory Group, D-22603 Hamburg, Germany}
\end{center}
\vskip 3.cm

\begin{center}
{\bf Abstract}
\end{center}
A summary of new experimental results and 
recent theoretical developments presented 
in the ``Spin Physics'' working group is given.  

\vspace{6.5cm}
\noindent
{\it Convener's summary of the Working Group IV ``Spin Physics'' 
of the  5th International Workshop on 
``Deep Inelastic Scattering and QCD'' 
(DIS '97), Chicago, Illinois, USA, April 14-18, 1997.}
\vfill

\setcounter{page}{0} 
\newpage

\title{Spin Physics}

\author{T.~Gehrmann}
\address{DESY, Theory Group, D-22603 Hamburg, Germany}

\maketitle

\begin{abstract}
A summary of new experimental results and 
recent theoretical developments presented 
in the ``Spin Physics'' working group is given.  
\end{abstract}

\section{Introduction}
One of the most intriguing, and yet unanswered, questions in 
particle physics is to find out how the spin of the proton is 
shared among its constituents, the quarks and gluons. The experimental 
observation~\cite{emc} that the contribution of quarks to the proton
spin is far smaller than naively expected in the Ellis--Jaffe sum 
rule~\cite{ej}, 
made by EMC 
nine years ago, initiated a huge experimental programme of spin 
structure function measurements at CERN, SLAC and DESY. All these 
experiments have confirmed the original EMC observation
with a continuously improving level of precision and provided 
new insights into the nucleon spin structure from the 
measurement of various 
other inclusive and semi-inclusive observables. Moreover, the EMC
result has motivated 
much theoretical 
work towards a better understanding of the nucleon's spin. 
The study of the 
spin structure of the nucleon has become by now
an important aspect of deep inelastic scattering, 
and one working group of the present workshop
was devoted to ``Spin Physics''. In this brief review, I will attempt to
summarize the results and 
developments discussed in this group. The choice of material presented here 
is necessarily restricted and can only give a first impression of the current
trends in spin physics.

\section{New experimental results}
\label{sec:exp}
Deep inelastic scattering off polarized targets is presently studied 
at three different experiments.
The SMC experiment~\cite{magnon,kabuss} in  the CERN
190~GeV polarized muon beam uses  a large polarized solid state target; this 
experiment has been operational from 1992-96 and has recently
finished data-taking. 
The major improvement in the 1996 run was the use of 
ammonia as target material which has a higher target dilution factor 
(fraction of polarizable protons in the target) 
than the previously used butanol. Of all polarized DIS experiments, 
SMC has the largest beam energy and therefore covers 
lower values of $x$ than its competitors.  

A series of SLAC experiments 
is studying spin structure functions with the polarized SLAC electron 
beam. The most recent experiments~\cite{meziani}
in this series were E154 and E155,
working at an electron beam energy of 48.3~GeV, compared to 
beam energies of $10-29$~GeV available to their predecessors. 
Moreover, the degree of beam polarization has been improved with respect 
to the earlier SLAC measurements. These experiments use a  
target cell with polarized $^3$He (E154) or polarized ammonia (E155). 
The E154 experiment, which was carried out early last year, has already 
published first results; E155 has just completed data-taking.     

The youngest competitor in polarized deep inelastic scattering is 
the HERMES experiment~\cite{stoesslein,schuler}
operating a polarized internal gas target
in the HERA 27.5~GeV positron beam, which is polarized naturally due to 
the Sokolov--Ternov effect. This experiment offers  a presently unique 
identification of particles
in hadronic final state of deep inelastic scattering 
and is therefore ideal for the measurement of semi-inclusive 
asymmetries.  

\subsection{Inclusive measurements and sum rules}
The study of the polarized structure functions $g_1(x,Q^2)$ and 
$g_2(x,Q^2)$, measured in inclusive lepton-nucleon scattering 
in the above experiments yields various different insights into the 
spin structure of the nucleon. The structure function $g_1$ in particular 
has a simple partonic interpretation as the charge weighted sum 
of the quark polarizations in the nucleon,
\begin{displaymath}
g_1(x,Q^2) = \frac{1}{2} \sum_{q,\bar q} e_q^2 \left[ q^\uparrow (x,Q^2)
- q^\downarrow (x,Q^2) \right]\; ,
\end{displaymath}
thus yielding information on the polarized parton distributions in the 
nucleon. Sum rules due to Bjorken~\cite{bj} and Ellis and Jaffe~\cite{ej} 
relate the first moment of $g_1$, 
\begin{displaymath}
\Gamma^{p,d,n}_1(Q^2) = \int_0^1\, g_1^{p,d,n} (x,Q^2)\, {\rm d} x\; ,  
\end{displaymath}
to the axial vector coupling constants measured in $\beta$-decays. 
The test 
of these sum rules is clearly one of the key issues in spin physics. It 
should however always be kept in mind that only the Bjorken sum rule 
is a rigid prediction of isospin symmetry, while the 
Ellis--Jaffe sum rule is based on a much weaker footing in the naive quark
parton model. 
\begin{figure}[b!] 
\centerline{\epsfig{file=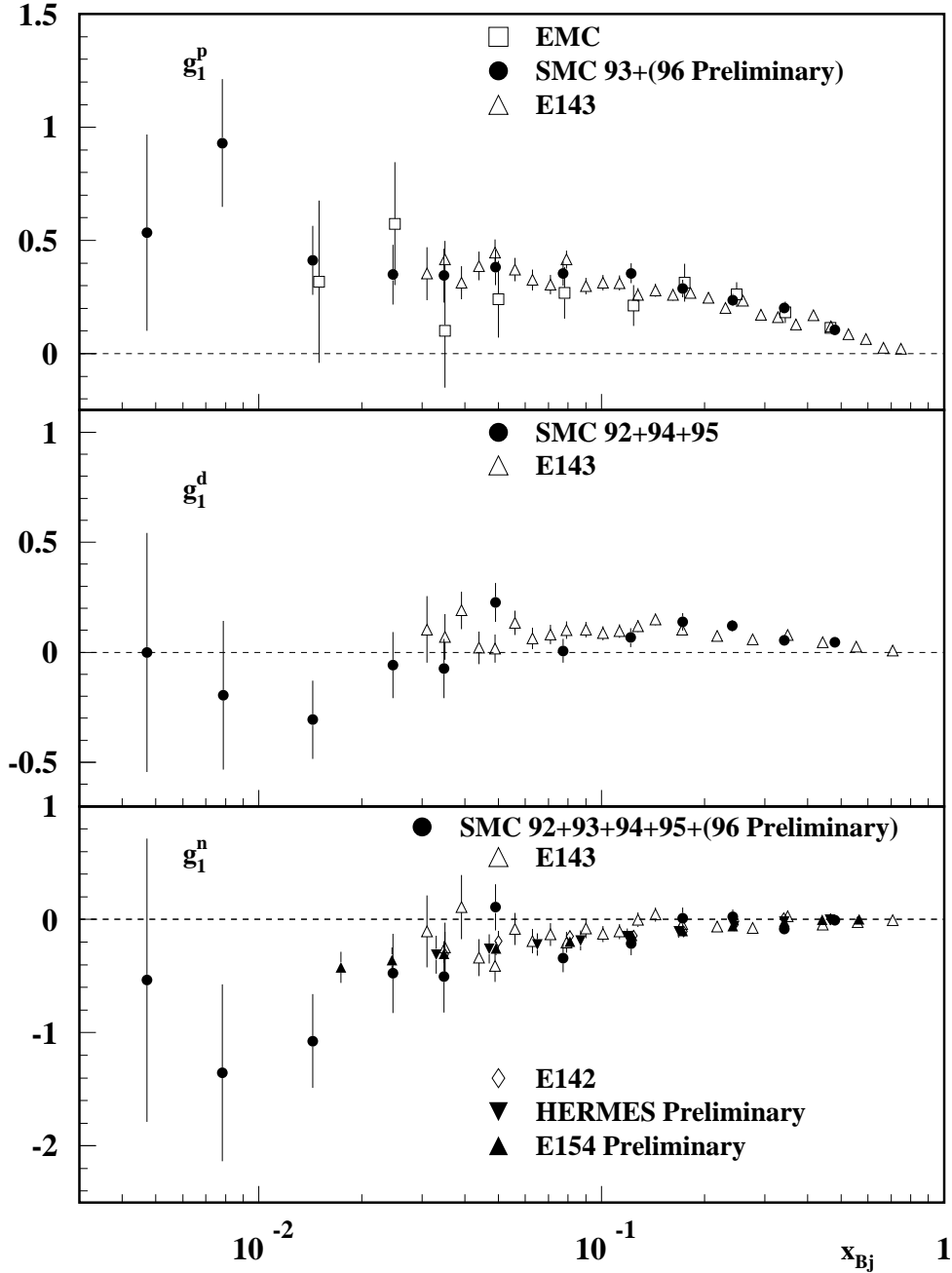,width=13cm}}
\vspace{10pt}
\caption{World data on the spin structure function \protect $g_1(x,Q^2)$ 
evolved  
to \protect $Q^2=5$~GeV\protect $^2$.}
\label{fig:g1}
\end{figure}

Based on the data taken in 1996, the SMC experiment has recently presented 
a new (preliminary)
measurement of the proton spin structure function 
$g_1^p(x,Q^2)$~\cite{magnon}. 
Compared with the earlier SMC measurement of $g_1^p(x,Q^2)$, statistical 
errors have now been reduced by a factor of 2.  One of the most 
striking results of this measurement is the behaviour of $g_1^p$ 
at small $x$. In contrast to earlier SMC results clearly
indicating  a  rise
of $g_1^p$, 
one finds this rise at small $x$ to be rather moderate now if all SMC 
proton data are combined, as can be seen in Figure~\ref{fig:g1}. It must 
however be pointed out that the $x$ values currently probed at 
polarized fixed target experiments are about two orders of magnitude larger 
than the $x$ values probed in unpolarized collisions at HERA. It is therefore 
at least doubtful that the kinematic region covered at SMC can yield 
conclusive information on the small-$x$ behaviour of the polarized structure 
functions. 

Using the 1996 proton data, SMC has moreover presented an improved 
measurement~\cite{magnon} of the Ellis--Jaffe proton
sum rule $\Gamma_1^p (Q^2 = 10~\mbox{GeV}^2)
= 0.149 \pm 0.012$, which is more than 1.5$\sigma$ below the 
prediction of Ellis and Jaffe.
The Bjorken sum rule 
is found to be 
$\Gamma_1^{p-n}(Q^2 = 10~\mbox{GeV}^2) = 0.209 \pm 0.026$, which is 
consistent with the predicted value of 0.187$\pm$0.002.

Both HERMES and E154 have measured the neutron spin structure function 
$g_1^n(x,Q^2)$ off polarized $^3$He-targets~\cite{meziani,stoesslein,slacg1},
the results are included in Figure~\ref{fig:g1}. Based on these recent 
data, both experiments have presented new determinations of the 
Ellis--Jaffe neutron sum rule, HERMES: $\Gamma_1^n (Q^2 = 2.5~\mbox{GeV}^2) 
= -0.037 \pm 0.013 \pm 0.005$ and E154: $\Gamma_1^n (Q^2 = 5~\mbox{GeV}^2) 
= -0.041 \pm 0.004 \pm 0.006$. These measurements are consistent 
with each other and significantly below the
value predicted by Ellis and Jaffe.

A compilation of all world data~\cite{emc,magnon,meziani,stoesslein,world}
on the structure function 
$g_1$ of the proton, deuteron and neutron, evolved to a 
common value of $Q^2=5\;\mbox{GeV}^2$ is shown in Figure~\ref{fig:g1}. 
 
All above measurements of the spin sum rules for $\Gamma_1^p$ and $\Gamma_1^n$
face two common problems: the evolution of all data points taken in the 
experiment to a common value of $Q^2$, where the sum rule is evaluated,  
and the extrapolation of $g_1$ between the lowest measured $x$ point and 
$x=0$. 

The evolution of all data points to a common value of $Q^2$ was up to 
very recently made by assuming that the 
structure function ratio  $A_1(x,Q^2)\sim g_1(x,Q^2)/F_1(x,Q^2)$ 
is independent of $Q^2$. This assumption  
is consistent with present experimental data on the $Q^2$--dependence of 
$A_1$, which cover however only a relatively narrow 
range in $Q^2$ for fixed $x$. Perturbative QCD 
on the other hand predicts
 a non-vanishing $Q^2$--dependence of $A_1$. An 
improved evolution 
procedure, incorporating the results of QCD analyses of the 
data on $g_1(x,Q^2)$, is now used --~at least 
as a cross check of the above method~-- in all recent experimental evaluations 
of the Ellis--Jaffe sum rule. This method will be discussed below in 
Section~\ref{sec:qcd}. 
The difference between the 
results obtained in both methods is
however still significantly smaller than the present statistical and 
systematical errors on the measured values of $\Gamma_1^{p,n}(Q^2)$.

The extrapolation of $g_1(x,Q^2)$ into the experimentally unmeasured 
small $x$ region has by now become one of the major sources of 
uncertainty in measurements of the spin sum rules. This extrapolation 
has up to now been performed assuming $g_1(x\to 0) \sim const$, which 
is motivated by Regge theory. This behaviour is however immediately
 broken by QCD evolution, which predicts $g_1(x)$ to rise at least 
logarithmically at small $x$. A study of the small $x$ behaviour of 
$g_1^n(x,Q^2)$ by E154~\cite{meziani} showed that the 
present data are consistent both with $g_1(x)=const$ and the extreme 
behaviour $g_1(x)=C/x^{0.9}$, indicating that the neutron spin 
structure at small $x$ is still largely unknown -- despite the 
experimental progress made on the neutron spin structure
in recent times. This 
induces consequently a non-quantifiable uncertainty on the spin sum 
rules arising from the small-$x$  region. 
\begin{figure}[t!] 
\centerline{\epsfig{file=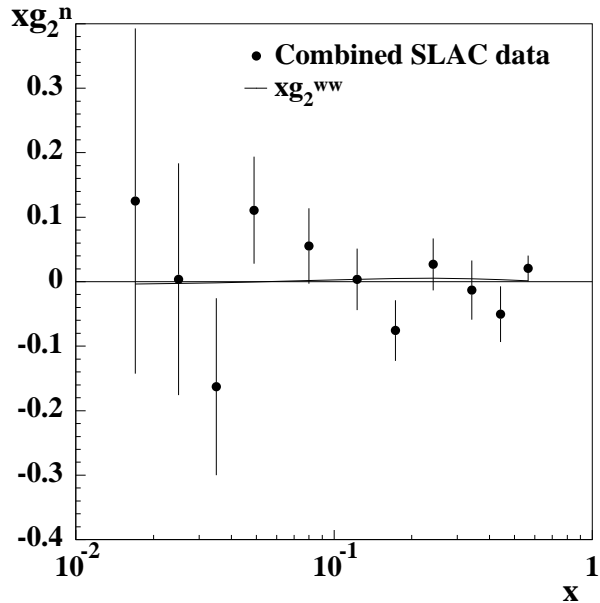,width=8cm}}
\vspace{10pt}
\caption{Combined SLAC data on the neutron spin structure function 
\protect $g^n_2(x)$.}
\label{fig:g2}
\end{figure}

Apart from the measurement of $g_1(x,Q^2)$ in deep inelastic scattering 
off longitudinally polarized targets, the SLAC experiments have as 
well measured~\cite{meziani,slacg2}
the second spin structure function $g_2(x,Q^2)$, 
which is accessible using transverse target polarization. The result
of all SLAC measurements of $g^n_2(x,Q^2)$ at $Q^2 = 3$~GeV$^2$ 
is shown in Figure~\ref{fig:g2}. This 
structure function has no simple partonic interpretation but
is of particular interest, as it can  
receive sizable contributions from twist-3 operators.
Using an exact integral relation~\cite{ww} for the twist-2 
contributions to $g_1$ and $g_2$, it is possible to predict the 
twist-2 content of $g_2$, indicated by the solid line in Figure~\ref{fig:g2}.  
Using this integral relation, it is moreover possible to 
measure~\cite{meziani,slacg2} the  
twist-3 matrix element $d_2^n=(-1.0\pm 1.5)\times 10^{-2}$ from 
the second moment of $g_2$. This measurement is however not yet accurate 
enough to discriminate between different theoretical 
 predictions 
obtained from QCD sum rules~\cite{qcdg2}, in the bag model~\cite{bagg2}
and from lattice calculations~\cite{latg2}.

\subsection{Semi-inclusive measurements}
The information on the individual quark and anti-quark
polarizations in the nucleon obtained from inclusive measurements of 
the structure function $g_1$ is naturally limited by the fact that 
two independent functions, $g_1^p$ and $g_1^n$, are insufficient to 
disentangle the five different distributions $\Delta u_v$, $\Delta d_v$,
$\Delta \bar u$, $\Delta \bar d$ and $\Delta s$. A possible way to 
access these individual distributions in deep inelastic scattering is
the study of semi-inclusive asymmetries, defined by  
a particular hadron observed in the final state. 
\begin{figure}[b!] 
\centerline{\epsfig{file=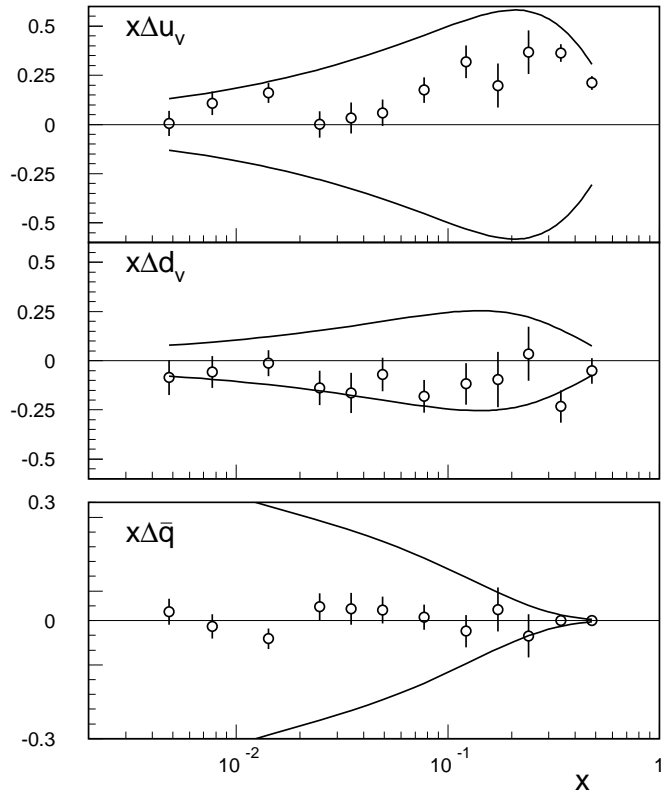,width=8.7cm}}
\vspace{10pt}
\caption{Polarized parton distributions obtained from the SMC analysis 
of semi-inclusive asymmetries (preliminary).
The corresponding unpolarized distributions 
are indicated by the solid lines.}
\label{fig:hadron}
\end{figure}

The first results on semi-inclusive asymmetries for positively and 
negatively charged hadrons were published already some time 
ago by SMC~\cite{smchadrons}.
Based now on the whole SMC data sample, an improved (preliminary) 
determination of $\Delta u_v$, $\Delta d_v$ 
and $\Delta \bar q$ from semi-inclusive asymmetries has been presented 
during the workshop~\cite{kabuss}, the results of this analysis are 
shown in Figure~\ref{fig:hadron}.
A measurement of semi-inclusive 
asymmetries for particular hadron species, required for a flavour 
decomposition of the light quark sea $\Delta \bar q$, is however 
not possible with the SMC apparatus, which does not have a dedicated 
hadron identification. 

First hadron results obtained at HERMES~\cite{schuler}, 
e.g.~the 
decay angle distribution in $\rho \to \pi^+\pi^-$ or the 
ratio of the unpolarized valence quark distributions $d_v/u_v$ demonstrate 
the potential of the HERMES apparatus for precision measurements of 
semi-inclusive asymmetries in polarized deep inelastic scattering.  
The first HERMES results on these can be expected in the near future. 

\section{Status of polarized parton distributions}
\label{sec:qcd}

\subsection{QCD analysis of polarized structure function data}
Using the recently calculated polarized two-loop splitting 
functions~\cite{twoloop}, it is now possible to perform consistent
next-to-leading order fits~\cite{gs,grsv,abfr,zyla}
of polarized parton distributions to the world data on $g_1(x,Q^2)$. 
There are various motivations for these fits: (i) the parton distributions 
are universal, process-independent features of the nucleon, their 
knowledge therefore enables the calculation of a variety of 
hard cross sections 
in polarized hadron--hadron collisions; (ii) the knowledge of 
the polarized parton distributions allows one to quantify the 
effects of QCD evolution on the structure function $g_1$, which is 
crucial for the comparison of data taken at different $Q^2$ and 
for the evaluation of spin sum rules (see above); (iii)
the resulting distributions can be compared with non-perturbative 
calculations (e.g.~in Lattice-QCD, see below). 

All QCD fits assume simple parametric forms for the initial distributions 
at some low scale $Q_0^2$, which are then evolved according to the 
DGLAP evolution equations~\cite{dglap} and fitted to the 
experimental data on the structure function $g_1$ at higher $Q^2$.
There is some ambiguity in the choice of factorization scheme at 
next-to-leading order. At present, two schemes are commonly used in 
the QCD fits: the well-known $\overline{{\rm MS}}$ scheme and the 
AB scheme, allowing for a non-zero gluonic contribution to the 
Ellis--Jaffe sum rule. Fits in both schemes yield equally 
good descriptions of the structure function data, as shown in~\cite{zyla}.
Two new QCD analyses of the world data on polarized structure functions 
(excluding the only recently released new SMC proton data)
were presented during the workshop~\cite{abfr,zyla}. 

\begin{figure}[t!] 
\centerline{\epsfig{file=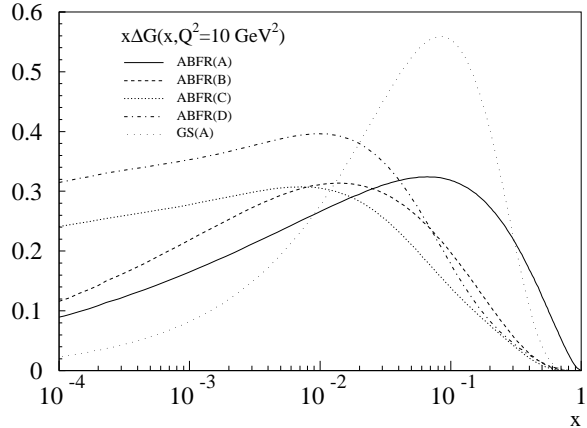,angle=-90,width=7.5cm}}
\vspace{10pt}
\caption{Different parameterizations of the polarized 
gluon distribution equally allowed by present data
from the recent ABFR analysis \protect\cite{abfr} and the earlier
GS analysis~\protect\cite{gs}.}
\label{fig:gluon}
\end{figure}

All QCD analyses of the polarized structure function data 
come to the common conclusion that only the valence quark 
polarization and the total sea quark polarization are 
well constrained by the experimental data. The precise 
flavour decomposition of the polarized quark sea and the 
polarization of gluons in the nucleon are at present largely 
unknown. The $Q^2$ evolution between the data-sets of SLAC and SMC,
taken at different beam energies, indicates however a positive 
overall gluon polarization $\Delta G(Q^2=5~\mbox{GeV}^2) \sim {\cal O}(2)$,
but leaves the $x$--dependence of
the corresponding distribution~$\Delta G(x,Q^2)$ rather unconstrained.
This is illustrated in Figure~\ref{fig:gluon}, showing four different 
parameterizations of $\Delta G(x,Q^2=10~\mbox{GeV}^2)$ obtained in the 
recent analysis of~\cite{abfr} together with one parameterization 
obtained earlier in~\cite{gs}. Although these parameterizations are 
in different schemes (AB~\cite{abfr} and $\overline{{\rm MS}}$~\cite{gs}), 
they 
are still comparable, as the corresponding scheme transformation 
leaves the polarized gluon distribution unaffected.
 All parameterizations shown in Figure~\ref{fig:gluon}
yield equally good
descriptions of the polarized structure function data. 
Finally, all fits yield consistent values for the first moment of the 
singlet axial vector current $a_0$, which can in the 
$\overline{{\rm MS}}$ scheme be identified with 
total quark 
contribution $\Delta \Sigma$ to the proton spin. The current value for
this quantity is 
$\Delta \Sigma^{\overline{{\rm MS}}} 
(Q^2=5~\mbox{GeV}^2) = 0.20 \pm 0.08$~\cite{zyla}. 

Using the results of these QCD fits, it is possible to 
evolve all data points of an experiment to a common value of $\langle Q^2
\rangle $,
as required for evaluations of the Ellis--Jaffe sum rule. Defining 
 the shift parameter 
\begin{displaymath}
\Delta g_1^{{\rm fit}} (x_i,Q^2_i,\langle Q^2 \rangle ) \equiv
g_1^{{\rm fit}} (x_i,Q^2_i) - g_1^{{\rm fit}} 
(x_i, \langle Q^2 \rangle )
\end{displaymath}
for each data point $(x_i,Q^2_i)$, the value of $g_1^{{\rm exp}} (x_i,
 \langle Q^2 \rangle )$ can be approximated by
\begin{displaymath}
g_1^{{\rm exp}} (x_i,\langle Q^2 \rangle ) = g_1^{{\rm exp}} (x_i,Q_i^2) 
- \Delta g_1^{{\rm fit}} (x_i,Q^2_i,\langle Q^2 \rangle ). 
\end{displaymath}
This procedure enables a consistent  estimate of the
systematic error induced by the evolution effects. The integral of 
$g_1(x,Q^2)$ over the $x$ range covered by experimental data obtained
with this method is usually consistent with the results obtained 
assuming $A_1$ to be independent of $Q^2$.

Believing that the present data are already sufficient to 
constrain the behaviour of the polarized quark and gluon distributions 
at small $x$, it is furthermore possible to predict the behaviour of 
the polarized structure function $g_1$ in the small $x$ region 
(disregarding potential effects
due to resummations~\cite{bar} of 
terms of ${\cal O} (\alpha_s^n \ln^{2n} x$)). The small $x$ 
extrapolations  
obtained from the fits in this way are systematically more 
singular~\cite{abfr,zyla}    
than the small $x$ extrapolations  from Regge theory, predicting
$g_1\sim const$ at small $x$.
In particular, both recent QCD analyses~\cite{abfr,zyla} found that  
the small $x$ contribution to $\Gamma_1^n$ shifts the central 
value of this sum rule by more than two standard deviations. 

The Bjorken sum rule is on the other hand less affected by the
uncertainties arising in the small $x$ region, in particular since 
it corresponds to a non-singlet combination of the polarized 
parton densities evolving independently of the polarized quark singlet 
and polarized gluon distribution. Using the small $x$ extrapolations 
motivated by the parton distribution fits, the Bjorken sum rule 
is found~\cite{abfr,zyla} 
to be within one standard deviation of the predicted value. 
This sum rule can finally be used for a determination 
of the nucleon axial vector coupling $g_A=1.19\pm 0.09$~\cite{abfr}, 
which is consistent with the value obtained from neutron 
$\beta$-decay $g_A=1.257\pm 0.003$. 
\begin{figure}[b!] 
\centerline{\epsfig{file=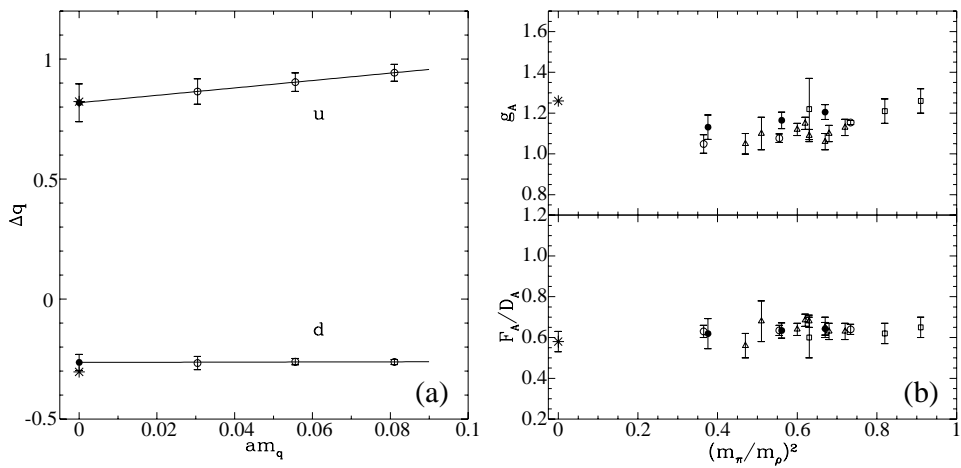,width=13.5cm}}
\vspace{10pt}
\caption{Lattice results on (a) the first moments of the polarized 
valence quark distributions compared to fit results (stars) and 
(b) the SU(3)\protect $_f$ axial vector coupling constants.}
\label{fig:lattice}
\end{figure}

\subsection{Lattice results}
The calculation of the lower moments of (unpolarized and polarized)
parton distribution functions 
in the nucleon has made considerable progress over the last two 
years~\cite{schierh}. The most recent improvement in this field is a
systematic procedure for the removal of all terms linear in the lattice 
spacing from the lattice observables, yielding a better extrapolation 
towards the continuum limit. First results obtained with this 
method were shown during the conference. These include 
improved lattice determinations 
of the  first moment of the 
polarized $u$ and $d$ valence quark distributions, the axial vector 
current $g_A$ (Bjorken sum rule) 
for the ratio of the hyperon decay constants $F/D$ (Ellis-Jaffe sum rule).
The interpolation of these results towards the continuum limit is 
shown in Figure~\ref{fig:lattice}. The results obtained for the 
polarized
valence quark distributions are in very good agreement with the results 
of recent fits to the structure function data~\cite{gs}. While the 
determination of the $F/D$ ratio is consistent with experimental 
data from hyperon decay, there appears to be a discrepancy of about 10\%
between the 
lattice result for $g_A$ and the value obtained from neutron $\beta$-decay. 
The source of this discrepancy is not understood at present, it might 
however indicate a failure of the naive parton model identification of 
this current on the lattice. 
 
\section{Theoretical developments}
\label{sec:theo}
\subsection{Progress in higher order corrections}
While QCD corrections~\cite{bjho,ejho}
 to the spin sum rules are already known to
${\cal O}(\alpha_s^3)$ (these corrections to the Ellis--Jaffe 
sum rule were calculated only recently and are discussed in~\cite{ejho}),
one was restricted to lowest order approximations in studies of 
most other spin-dependent observables up to now.
Only the calculation~\cite{twoloop} of the space-like polarized two 
loop splitting functions $\Delta P_{ji}(x)$, crucial ingredients for a 
determination of polarized parton distributions at next-to-leading 
order~\cite{gs,grsv,abfr,zyla}, enables now consistent studies of 
polarized observables beyond leading order. Several new 
results on higher order corrections to spin-dependent 
processes were presented during the workshop. 

Much information on the unpolarized sea quark distributions in the 
nucleon was gained from experiments on lepton pair production at fixed 
target energies (the Drell-Yan process, for recent results see
e.g.~\cite{dynew}). Moreover, the production of vector bosons 
at collider energies is mediated by the same process. Given the 
importance of higher order corrections to the unpolarized Drell--Yan 
cross section, the knowledge of QCD corrections to the polarized Drell--Yan
process will be crucial for a reliable interpretation of future data 
on vector boson production at RHIC. The complete 
${\cal O}(\alpha_s)$-corrections to the polarized Drell--Yan cross section
as function of the rapidity $y$ and the Feynman-parameter $x_F$ have been 
calculated recently~\cite{dypol}. These corrections turn out to 
be similar to the corrections in the unpolarized case and hence 
numerically sizable even at collider energies. 
Furthermore, some progress towards the calculation 
of the Drell--Yan cross section at  ${\cal O}(\alpha_s^2)$ has been
made with the calculation of the non--singlet contributions to the 
Drell--Yan cross section for non-zero transverse momentum at this 
order~\cite{gordon}.

Finally, the time-like polarized splitting functions
$\Delta P_{ij}(z)$, related 
to the spin transfer in fragmentation processes have been 
recently derived~\cite{ptime,fsv}
by use of analytic continuation relations applied to their space-like 
counterparts. First applications to
 polarized $\Lambda$-production
will be discussed in the following. 

\subsection{Spin-transfer in semi-inclusive reactions}
\begin{figure}[t!] 
\centerline{\epsfig{file=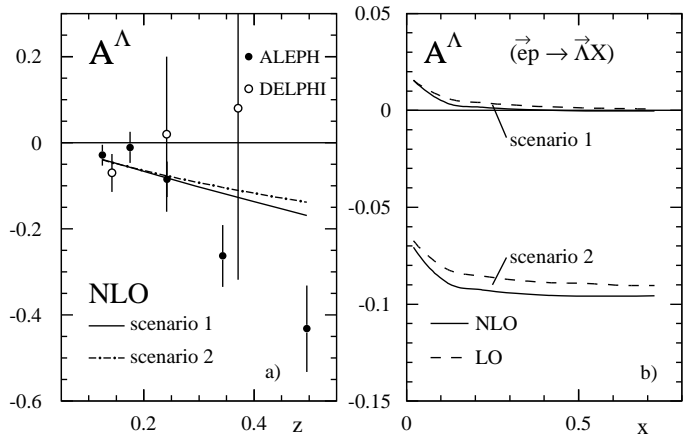,width=9cm}}
\vspace{10pt}
\caption{\protect $\Lambda$ 
polarization in \protect $Z$ boson decays (a) and semi-inclusive
DIS with polarized leptons at HERMES
(b) for different scenarios of the spin 
transfer in \protect $\Lambda$ fragmentation.}
\label{fig:lambda}
\end{figure}
The polarized parton distributions describe  
the probability of finding a parton of a particular species having its 
spin aligned or anti-aligned with the spin of the nucleon.
Correspondingly, one can define polarized fragmentation functions 
parameterizing the probability of a polarized parton fragmenting into a
hadron with spin aligned or anti-aligned to the parent parton spin. 
These polarized fragmentation functions are however experimentally 
only very hard to 
access for most hadrons, as they require the measurement 
of the spin state of a final state particle. Such a 
measurement is in 
practice only feasible for particles with dominant parity violating 
decay modes such as the $\Lambda$ baryon. 
First studies~\cite{fsv,bralambda} 
on the polarized 
fragmentation functions into $\Lambda$'s have been carried out 
recently. These studies consider two possible 
scenarios for the spin transfer to the $\Lambda$: a naive 
approach~\cite{lamnaiv} in 
which the $\Lambda$ spin is carried only by the $s$ quark and an 
approach due to~\cite{bi}, where the spin is shared among the $u,d,s$
quarks in the $\Lambda$. It has been shown in~\cite{fsv} that 
a fit to LEP data~\cite{lamdata} only is insufficient to discriminate the 
two scenarios, whereas a clear distinction would be possible in 
semi-inclusive DIS 
with a polarized lepton beam onto an unpolarized proton target, as can be
seen in Figure~\ref{fig:lambda}.

\subsection{Instanton calculations}
Several attempts to calculate contributions to deep inelastic 
structure functions induced by instantons have been made recently
(see e.g.~\cite{insrev} for a review). There are two substantially 
different approaches to the phenomenology of instanton effects 
in deep inelastic scattering. Several predictions for 
instanton contributions to the polarized parton distributions
obtained in the 
instanton liquid model~\cite{insliq} have been presented during this 
workshop~\cite{ko}. This approach predicts a large and negative 
contribution to the polarized gluon distribution and an approximate 
relation $\Delta \bar d (x) \approx -2\Delta \bar u(x)$. Factorization 
properties and infrared behaviour within this model,
based on an instanton lagrangian for fixed instanton radius, 
are however not yet clear at 
present. A substantially
different approach is the systematic treatment 
within instanton perturbation theory~\cite{insgas}. 
This approach
is free of infrared problems, as a dynamical cut-off on the instanton
size is provided by the typical hard scale of the scattering process.
Predictions obtained with this method concern mainly 
the structure of the multi-particle final state in deep inelastic 
scattering~\cite{insrev}, 
and predictions for polarized observables are 
not yet available.

\subsection{$g_2(x,Q^2)$ at small $x$}

Analytical calculations~\cite{bar} of the asymptotically dominant 
logarithmic contributions $(\alpha_s \ln^2 (1/x))^n$
to the polarized structure function $g_1(x,Q^2)$, 
based on the resummation of soft gluon contributions 
in an infra-red evolution equation~\cite{kir}, have now been available 
for some time, and contributions to the chirally-odd polarized structure 
function $h_1$ have been calculated recently~\cite{h1}.
 Based on a similar approach, the non-singlet 
contribution to the polarized structure function $g_2(x,Q^2)$ has now 
been studied in~\cite{ermo}. This calculation yielded a simple relation 
\begin{displaymath}
g^{{\rm n.s.}}_2\sim \frac{\partial g^{{\rm n.s.}}_1}{\partial \ln \alpha_s}
\; ,
\end{displaymath}
with $\alpha_s$ corresponding to ladder gluon contributions only, between
$g^{{\rm n.s.}}_1$ and $g^{{\rm n.s.}}_2$, 
suggesting both to have the same small-$x$
behaviour. It is however not clear at present, to which extent the above 
result is consistent with the Wandzura-Wilczek sum rule~\cite{ww} and 
whether contributions to $g_2$ from twist-3 operators become important 
at small $x$.

\subsection{Sum rules in twist-2 and twist-3}
A variety of different sum rules and integral relations
for the polarized structure functions
appearing in the deep inelastic scattering cross sections for 
neutral and charged current exchange, mostly only valid for the 
twist-2 contributions, have been derived in the past. 
Motivated by several apparent discrepancies in the literature, 
a detailed re-investigation of the validity of sum rules and 
integral relations in twist-2 in
polarized deep inelastic scattering has been performed~\cite{sumrule} 
over the past year. This study, consistently 
carried out in the operator product expansion and cross-checked in 
the covariant parton model, confirmed only some of the earlier results while 
disproving others. Moreover, several new relations involving the 
polarized charged current structure functions have been derived 
in~\cite{sumrule}, 
yielding finally a self-consistent and complete picture of the twist-2 
contributions to polarized deep inelastic scattering. 
Moreover, higher twist contributions
 to the polarized structure functions, in particular the presumably 
non-negligible twist-3 contributions to $g_2$ and $g_3$, 
can be extracted by using exact relations for their twist-2 content. 
A new relation between the twist-3 contributions to $g_2$ and 
$g_3$ has been derived in~\cite{sumrule}.

Finally, the work of~\cite{sumrule} could verify that a sum rule
for the valence content of $g_1$ and $g_2$,
\begin{displaymath}
\int_0^1 {\rm d} x \; x \left[ g_1^V(x) + 2 g_2^V (x) \right] =0 ,
\end{displaymath}
 which was derived earlier in~\cite{efr} is consistent with 
the operator product expansion in massless QCD, although it cannot be 
explicitly proven in this approach. 

\subsection{Exclusive reactions}
The total spin of the nucleon does not only receive contributions 
from the spin carried by its constituents, quarks ($\Delta \Sigma$)
and gluons ($\Delta G$), but as well from the orbital angular momentum of 
quarks $L_{\Sigma}$ and gluons $L_G$:
\begin{displaymath}
\frac{1}{2} = \frac{1}{2} \Delta \Sigma + \Delta G + L_{\Sigma} + L_G.
\end{displaymath}
Only the partonic spin contributions $\Delta \Sigma$ and $\Delta G$
to the nucleon spin can be accessed in inclusive or semi-inclusive 
reactions studied at present and future spin experiments. Up to very 
recently, no experimental observable was known to access the 
orbital angular momentum of the partons. It has been proposed 
recently~\cite{ji},
that the total quark angular momentum $J_{\Sigma}
= \Delta \Sigma/2 + L_{\Sigma}$
could be related the form factors accessible in the (unpolarized)
exclusive reaction
$\gamma^{\star}p 
\to p \gamma$ (Deeply Virtual Compton Scattering, DVCS) at 
large virtualities of the incoming photon and zero momentum transfer to 
the proton. First numerical studies of this reaction~\cite{guichon}
indicate that these form factors may be measurable at the COMPASS 
experiment discussed below, while the DVCS cross section at HERMES 
is concealed by a large QED background. 

Although the identification of the total quark angular momentum with 
exclusive form factors is not undisputed, it has triggered a large 
interest in the perturbative description of exclusive reactions involving 
a large momentum transfer. It has already been proposed about ten 
years ago~\cite{robaschik} that exclusive reactions in $ep$ scattering at
large momentum transfers $Q^2$ and small proton deflection $t$
could be described by non-forward parton distributions, being 
hybrids of ordinary parton distributions~\cite{dglap} and 
distribution amplitudes~\cite{bl} describing e.g.~meson formation from
two parton initial states. These non-forward parton distributions~\cite{rady},
functions of two scalar variables describing all hard partonic momenta 
in the exclusive reaction,
obey evolution equations which can be reduced to the common DGLAP~\cite{dglap}
or BL~\cite{bl} evolution equations for parton distributions or 
distribution amplitudes by respective integration. Moreover,
various integral relations between non-forward parton distributions 
and elastic form factors or parton distributions can be derived~\cite{rady}.
The unpolarized
one-loop evolution kernels for non-forward parton distributions have already 
been known for some time~\cite{robaschik},
and the full spin dependence of these 
kernels was derived recently~\cite{robnew}.

\section{Future experiments}
\label{sec:futexp}
The completion of the SLAC programme and the SMC experiment marks 
the end of the ``second generation'' of dedicated polarized structure 
function measurements. 
Only the HERMES experiment will continue 
to provide new data on $g_1$ and $g_2$, together with 
the first precision measurements 
of semi-inclusive asymmetries.

Our knowledge on the spin structure of the nucleon is at present 
largely based on the structure function measurements discussed 
above and therefore inevitably incomplete. 
The study of the unpolarized proton structure has shown that 
information obtained from structure functions 
alone is insufficient for an unambiguous determination of all different 
parton distribution functions in the nucleon. Only the combination of 
various observables from lepton-nucleon and nucleon-nucleon 
collisions in a global fit enables the extraction of quark distributions 
of different flavour and of the gluon distribution. Numerous
experiments devoted to the study of such complementary observables 
are presently  constructed, others are proposed or under 
discussion. 

Two experimental projects probing the spin structure of the 
nucleon were presented in more detail during the workshop: the 
recently approved COMPASS experiment~\cite{compass,bravar} at CERN and the 
operation of the HERA collider with a polarized proton 
beam~\cite{heraspin,deroeck}, 
which is one possible option for the mid-term future of HERA and
currently under study. The physics prospects of these two projects 
will be discussed in more detail below.

Another major project in spin physics is 
the polarized proton programme~\cite{rhic} at the
RHIC collider at BNL,  offering the possibility of studying 
polarized proton-proton collisions at centre-of-mass energies 
varying between 60~GeV and 500~GeV. This programme 
is expected to start in the year 2001 and will cover the broad 
range of physics observables accessible in hadronic collisions. 

Among the various spin experiments discussed at the moment are a
measurement of the polarized neutron structure function 
$g_1^n$ at large $x$ at CEBAF~\cite{meziani} 
and an upgrade of the HERMES spectrometer to enable 
measurements of open charm production~\cite{schuler}.

\subsection{COMPASS at CERN}
The COMPASS experiment~\cite{compass,bravar}
in the CERN polarized muon beam 
will use a newly built hadron and muon spectrometer to study 
inclusive and semi-inclusive scattering off a polarized nucleon target. 
It will start data taking in the year 2000. 

The key process studied at COMPASS is the production of 
open charm, induced by quasi-real photons. This process is mediated
by photon-gluon fusion and hence 
provides a direct probe of the polarized gluon distribution in the 
nucleon. Detailed simulations of the detection of open charm 
from reconstructed $D\to K \pi$ decays have shown~\cite{bravar} 
that COMPASS will be
able to measure the ratio of polarized to unpolarized gluon distribution
$\Delta G(x,Q^2)/G(x,Q^2)$ within an accuracy of $\pm$0.11 for $x\approx 0.1$ 
and $Q^2\approx 4 m_c^2$. This accuracy can be further improved by 
considering other decay channels of the $D$ meson, such studies  
are in progress.   

Another potential probe of the polarized gluon distribution at COMPASS
could be the production of oppositely charged hadron pairs
back-to-back at large transverse momentum. This process is currently
under study, and first results look promising~\cite{bravar}.

The COMPASS physics programme covers moreover improved measurements 
of the polarized structure functions $g_1$ and $g_2$ and of 
semi-inclusive asymmetries. Furthermore, studies of transversity 
distributions which are only 
accessible in semi-inclusive deep inelastic scattering
will be carried out. Finally, COMPASS will be able to study the 
spin transfer in semi-inclusive $\Lambda$ production discussed 
above~\cite{fsv,bralambda}.

\subsection{Future prospects for spin physics at HERA}
The commissioning of the HERA electron-proton collider five 
years ago opened up a completely new kinematical domain 
in deep inelastic scattering, and the two HERA experiments have provided 
a multitude of new insights into the structure of the 
proton and the photon since then. It is therefore only natural to 
assume that the operation of HERA with polarized proton and electron beams 
could add vital new information to our picture of the spin structure of the 
nucleon. The technical feasibility and physics prospects of this 
project have been investigated for the first time in a working 
group of last year's ``Future Physics at HERA'' workshop~\cite{heraspin}.
More detailed studies for several key observables are presently 
carried out in an ongoing workshop on ``Physics with Polarized 
Protons at HERA''~\cite{deroeck}. 

The HERA electron beam is polarized naturally due to the Sokolov--Ternov
effect, and stable electron polarization can be maintained over the 
whole beam lifetime. 
The polarization of the proton beam is on the other hand 
a major challenge in accelerator 
physics, since this requires the acceleration of polarized 
protons from low energies. The proton polarization has to be monitored and 
maintained during in the  whole chain of DESY pre-accelerators, which 
requires several major modifications in the beam optics. Despite the 
complexity of this undertaking, it appears to be technically feasible that 
a polarized proton beam could be operated at DESY in the mid-term 
future~\cite{desyspin}. 

The physics programme at a polarized HERA collider would be a natural 
continuation of the present unpolarized programme. 
 A measurement of 
the structure function $g_1^p$ will considerably reduce the 
uncertainty~\cite{abfr,zyla} on the Ellis--Jaffe sum rule arising from 
the extrapolation towards $x\to 0$ and provide information on the 
yet unknown behaviour of the polarized parton distributions at 
small $x$. A polarized HERA collider would be a unique place to study  
the weak polarized structure functions~\cite{weaksf} from the 
charged current cross section at large $Q^2$ and to explore the 
spin structure of photon from photoproduction of jets and heavy 
quarks~\cite{herapho}. Finally, a competitive measurement of the 
polarized gluon distribution could be obtained from the  
2+1 jet rate in deep inelastic scattering~\cite{disjet}. 

\section{Summary and outlook}
New measurements of the polarized structure function $g_1$, 
which have become available in the past year, have considerably 
improved on the precision of earlier data. 
However, these new measurements have as well raised new questions. 
The QCD analysis of the new data, which has become 
a standard procedure in 
all experimental studies by now, shows that the uncertainty 
on spin sum rules arising from the extrapolation of structure 
functions into the small $x$ region is far larger than previously assumed 
on the basis of Regge theory. These analyses  
illustrate moreover 
that the inclusive structure function alone is insufficient 
for a precise determination of the polarized gluon distribution from 
scaling violations and for a flavour decomposition of the polarized 
quark sea. These two aspects of the nucleon's spin structure 
will only become accessible in dedicated measurements at future 
spin experiments.  

The first measurements of the 
second spin structure function $g_2$ are now becoming available. With 
improving precision of the experimental data expected in the near future,
this structure function 
will become an important laboratory to access the twist-3 
component of the nucleon's spin structure. 

Much theoretical work has been devoted to various aspects of spin physics 
over the past year.  Several advancements in higher order 
corrections to spin observables and progress towards a 
perturbative description of exclusive reactions at large momentum 
transfer are among the theoretical highlights reported during the 
workshop.

A variety of new experimental information on the spin structure 
of the nucleon can be expected in the next years. The HERMES and E155
experiments will yield improved measurements of the inclusive 
structure functions $g_1$ and $g_2$, and HERMES will provide the 
first precision measurements of semi-inclusive asymmetries. The recently 
approved COMPASS experiment will start data-taking three years from 
now and will presumably yield the first direct measurement of the 
polarized gluon distribution from open charm photoproduction. A year 
later, the polarized proton programme at RHIC is expected to start operation. 
Finally, the polarization of the HERA proton beam, which is 
presently under discussion, would open up 
a completely new kinematic domain in the study of the spin structure 
of the nucleon and provide a multitude of new spin observables which are 
unique to HERA.   

\section*{Acknowledgements}

I am very grateful to   
Jos\'{e} Repond and his team
for the organization of such a very lively and exciting workshop,
and for the help and support I received from them in preparing this 
summary talk. It would not have been possible for me to summarize 
the broad spectrum of new results presented in our working group  
without the help from my fellow conveners Emlyn Hughes and Klaus Rith,
who shared their vast knowledge on the experimental aspects of 
spin physics with me and encouraged me to give this summary.
Finally, it is a pleasure to thank all participants who made 
the ``Spin Physics'' working group so interesting and successful and 
provided me with much of the material presented in this summary.

\end{document}